\documentclass{article}
\usepackage{amssymb,amsmath,cite,bm}
\def\T{\mathrm{T}}
\def\D{\mathrm{D}}

\def\Df{\mathbf{D}}
\def\M{\mathrm{M}}
\def\d{\mathrm{d}}
\def\r{\mathrm{r}}
\def\ri{\mathrm{i}}

\def\F{\mathcal{F}}
\def\Var{\mathrm{Var}}
\begin{document}

\begin{titlepage}
\begin{center}
{\large \textbf{Lack of anomalous diffusion in linear translationally-invariant systems
determined by only one initial condition}}

\vspace{2\baselineskip}
{\sffamily Mohammad~Khorrami~\footnote{e-mail: mamwad@mailaps.org},
Ahmad~Shariati~\footnote{e-mail: shariati@mailaps.org},
Amir~Aghamohammadi~\footnote{e-mail: mohamadi@alzahra.ac.ir}, \\ \&
Amir~H.~Fatollahi~\footnote{e-mail: ahfatol@gmail.com, tel/fax: ++98-21-88613937.}}

\vspace{2\baselineskip}
{\it Department of Physics, Alzahra University, Tehran 19938-93973, Iran}
\end{center}
\vspace{2\baselineskip}

\begin{abstract}
\noindent It is shown that as far as the linear diffusion equation meets
both time- and space- translational invariance, the time dependence
of a moment of degree $\alpha$ is a polynomial of degree at most
equal to $\alpha$, while all connected moments are at most linear functions
of time. As a special case, the variance is an at most linear function of time.
\end{abstract}
\vspace{\baselineskip}
\\[\baselineskip]
\end{titlepage}
\newpage
\section{Introduction}
Anomalous diffusion has found a special place in both observational and
theoretical sides of an extensive list of disciplines including physics,
chemistry, biology, economy, engineering, geology, meteorology and astrophysics
\cite{meka2000,hughes,bouch,meka2004,vlahos}.
This phenomenon is characterized by the long-time behavior of the variance of
some density behaving like
\begin{equation}
\Var(X)=\langle X^2 \rangle - \langle X \rangle ^2 \to C~ t^\alpha,
\qquad t\to\infty
\end{equation}
with $\alpha\neq 1$. For ordinary diffusion $\alpha=1$.

There has been extensive efforts to model the anomalous diffusion,
among them the continuous time random walks (CTRW)
\cite{meka2000,hughes,bouch,blu1986,shles}. Interestingly,
the Fourier-Laplace transform of the CTRW's equation may generate, among
other forms, the fractional diffusion equation, which has the form of an
ordinary diffusion equation but with non-integer order of derivatives
of space or time \cite{vlahos}. In recent years, great interest has
been attracted to the  study of the fractional dynamical equations
to model the anomalous diffusion behavior
\cite{barkai2001,hilfer2000,meka2000,meka2001}.

There have been efforts in other directions to explore in what situations
one may get the anomalous diffusion behavior.
In \cite{escande}, it was shown that the ordinary Fokker-Planck equation,
which is local, can give anomalous diffusive behavior in situations in which
there is a non-zero drift velocity. The cases with zero drift velocity are
explored in \cite{kl1987}, and it is shown that very particular forms of temporal and
spatial dependent diffusion coefficients can generate the anomalous diffusive
behavior. It has also been shown that particular non-linear diffusion
equations, which are like the ordinary ones in form, can produce the anomalous
diffusion \cite{lenzi}.

The aim of the present work is to show that systems satisfying the conditions
of linearity, both space and time translational invariance, and the property
that the state of the system at one time determines uniquely the state of
the system at later times, do not show anomalous diffusion.
Of course it does not mean that the lack of any of the above conditions
necessarily results in anomalous diffusion. Perhaps the most important
implication of this result is that it is not the appearance of
fractional derivatives, but some inhomogeneity hidden in the definitions
of some versions of such derivatives which result in anomalous diffusion.
In fact, there are versions of fractional derivative that commutate with
translation operators (an explicit example will be given), and it will be
shown that evolutions based on those do not result in anomalous diffusion.

The scheme of the
paper is the following. In section 2 the relation of time translation symmetry
and fractional time derivative is studied. Section 3 is on the solution of
fractional time derivative differential equations. In section 4 the relation
of moments corresponding to a distribution, and the Fourier transform of
that distribution are discussed. In section 5 the time evolution of a
distribution is discussed which satisfies a linear equation enjoying time
and space translation symmetry, and the property that the state of the system
at one time determines uniquely the state of the system at later times. It
is shown that the connected moments corresponding to such distributions
are at most linear functions of time. In section 6, an example is studied
which lacks time translation invariance and does exhibit anomalous diffusion.
The origin of this anomalous diffusion is then traced back (by explicit
calculation) to the lack of time translation invariance. Section 7 is devoted to
the concluding remarks.
\section{Time translation symmetry and the fractional time derivative}
There are different versions of fractional derivative. Demanding that
the fractional (time) derivative be commuting with time translation,
however, makes the fractional derivative almost unique. By almost it is
meant that as long as a specific branch of complex power function is chosen.
Denote by $\T_0(a)$ the operator which translates in time by $a$:
\begin{equation}\label{02.2}
[\T_0(a)\,\psi]:=\,\psi(t-a),
\end{equation}
and by $\D_0$ the time derivative. The generator of the group of
time translations is $(-\D_0)$:
\begin{equation}\label{02.3}
\T_0(a)=\exp(-a\,\D_0).
\end{equation}
Any operator which commutes with all time translations (in fact with
all time translations with the translation parameter in an interval
containing more than one point) commutes with the time derivative as well.
The converse is also obviously true. So the lack of time translation 
invariance of an operator (including a fractional time derivative),  
is equivalent to the non-commutativity of that operator with the
time derivative.
As the time derivative is not degenerate, one then concludes that any
such operator is a function of the time derivative. Denoting such an
operator with $\mathcal{O}$, there is a function $f$ such that
\begin{equation}\label{02.4}
\mathcal{O}=f(\D_0).
\end{equation}
The eigenvectors of $\D_0$ are exponentials. Denoting by $g(\centerdot)$
a function the value of which for the variable $t$ is $g(t)$, one has
\begin{equation}\label{02.5}
\D_0\,[\exp(s\,\centerdot)]=s\,[\exp(s\,\centerdot)],
\end{equation}
so,
\begin{equation}\label{02.6}
\mathcal{O}\,[\exp(s\,\centerdot)]=f(s)\,[\exp(s\,\centerdot)].
\end{equation}
To define the fractional time derivative, one simply takes $f$
to be a power function. But then, a power function with a
fractional exponent is not unique. To characterize it completely, one
should specify, in addition to the value of the exponent, the branch
cut in the complex plain. So is the case for the fractional derivative.

An example of a fractional time derivative defined in the literature,
which is \emph{not} commuting with time translation is the Caputo fractional
derivative $(\sideset{_b^\mathrm{C}}{_{0\,\beta}}{\mathop{\D}})$ \cite{frac1}:
\begin{equation}\label{02.7}
(\sideset{_b^\mathrm{C}}{_{0\,\beta}}{\mathop{\D}}\,\psi)(t)
:=\frac{1}{\Gamma(n-\beta)}\,\int_b^t\d\tau\;(t-\tau)^{n-\beta-1}\,\psi^{(n)}(\tau),\qquad n-1\leq\beta<n.
\end{equation}
It is seen that
\begin{equation}\label{02.8}
\T_a\,\sideset{_b^\mathrm{C}}{_{0\,\beta}}{\mathop{\D}}=
\sideset{_{b+a}^\mathrm{\quad C}}{_{0\,\beta}}{\mathop{\D}}\,\T_a.
\end{equation}
Another example of a fractional time derivative widely used in the literature,
which is \emph{not} commuting with time translation is the Riemann-Liouville 
fractional derivative $(\sideset{_{\;\;b}^\mathrm{RL}}{_{0\,\beta}}{\mathop{\D}})$ \cite{frac2}:
\begin{equation}\label{020.9}
(\sideset{_{\;\;b}^\mathrm{RL}}{_{0\,\beta}}{\mathop{\D}}\,\psi)(t)
:=\frac{1}{\Gamma(n-\beta)}\,\frac{\d^n}{\d t^n}\,\int_b^t\d\tau\;(t-\tau)^{n-\beta-1}\,\psi(\tau),\qquad n-1\leq\beta<n.
\end{equation}
It is again seen that
\begin{equation}\label{020.10}
\T_a\,\sideset{_{\;\;b}^\mathrm{RL}}{_{0\,\beta}}{\mathop{\D}}=
\sideset{_{b+a}^\mathrm{\;\;RL}}{_{0\,\beta}}{\mathop{\D}}\,\T_a.
\end{equation}

One example of a fractional time derivative defined in the literature,
which is commuting with time translation is the Weyl fractional
derivative $(\sideset{^\mathrm{W}}{_{0\,\beta}}{\mathop{\D}})$:
\begin{align}\label{02.9}
(\sideset{^\mathrm{W}}{_{0\,\beta}}{\mathop{\D}}\,\psi)(t):=\;&
\frac{(-1)^{m+1}}{\Gamma[m+1-\r(\beta)]}\,
\frac{\d^{[\beta]+1}}{\d t^{[\beta]+1}}\,\int_t^\infty\d\tau\;
(\tau-t)^{m-\r(\beta)}\,\psi^{(m)}(\tau),\nonumber\\
=\;&\frac{(-1)^{m+1}}{\Gamma[m+1-\r(\beta)]}\,
\frac{\d^{[\beta]+1}}{\d t^{[\beta]+1}}\,\int_0^\infty\d\tau\;
\tau^{m-\r(\beta)}\,\psi^{(m)}(\tau+t),
\end{align}
where $m$ is any nonnegative integer (the right hand side does not depend
on it, as long as the integrals are convergent), $[\beta]$ is
the largest integer not exceeding $\beta$, and
$\r(\beta)$ is the fractional part of $\beta$:
\begin{equation}\label{02.10}
\r(\beta):=\beta-[\beta].
\end{equation}
Direct calculation shows that if $s$ is a complex number of nonpositive
real part, then
\begin{equation}\label{02.11}
\sideset{^\mathrm{W}}{_{0\,\beta}}{\mathop{\D}}\,[\exp(s\,\centerdot)]=
\exp[-\ri\,\pi\,\r(\beta)]\,s^\beta\,[\exp(s\,\centerdot)],
\end{equation}
so
\begin{equation}\label{02.12}
\sideset{^\mathrm{W}}{_{0\,\beta}}{\mathop{\D}}=\exp[-\ri\,\pi\,\r(\beta)]\,\D_0^\beta.
\end{equation}
where the branch cut for the power function of exponent $\beta$ has been taken in the right
half plane, for example on the positive real semi axis.
\section{Solution to fractional time differential equations}
Consider the equation
\begin{equation}\label{02.13}
[f(\D_0)]\,\psi=0.
\end{equation}
A solution to this equation is
\begin{equation}\label{02.14}
\psi(t)=\oint_C\d s\;\frac{g(s)\,\exp(s\,t)}{f(s)},
\end{equation}
where $C$ is an arbitrary contour in the complex plane, and $g$ is
an analytic function. It is seen that with $\psi$ like (\ref{02.14}),
the left hand side of (\ref{02.13}) is an integral over $C$ of
the product $[g(s)\,\exp(s\,t)]$. If $g$ is analytic, then the
product is an analytic function of $s$ and the left hand side of
(\ref{02.13}) vanishes.

If $f$ is a polynomial of degree $n$, then the
singularities of the integrand in the right hand side of (\ref{02.14})
consist of only poles, which are the zeros of $f$. In that case
only the values of $g$ and some of its derivatives at those zeros
of $f$ which are encircled by $C$ enter the integral in the right hand side
of (\ref{02.14}). To be more specific, corresponding to a zero
$s_j$ only the derivatives of $g$ at $s_j$ of order less than $d_j$
enter, where $d_j$ is the degeneracy of $s_j$. The most general
solution to (\ref{02.13}) is then
\begin{equation}\label{02.15}
\psi(t)=\sum_{j=1}^k\sum_{\ell=1}^{d_j}a_{j\,\ell}\,t^{\ell-1}\,\exp(s_j\,t),
\end{equation}
where $a_{j\,\ell}$'s are arbitrary. One of course has
\begin{equation}\label{02.16}
\sum_{j=1}^k d_j=n.
\end{equation}
So the most general solution contains exactly $n$ arbitrary constants.

If $f$ is not a polynomial, then other sorts of singularities (branch cuts
and fundamental singularities) may arise as well. It may happen then, that
the general solution to (\ref{02.13}) contains an infinite number of arbitrary
constants.

Now consider equation (\ref{02.13}) together with an initial value condition
\begin{equation}\label{02.17}
\psi(t_0)=\psi_0,
\end{equation}
and the boundary condition at infinity
\begin{equation}\label{02.18}
|\psi(t)|<M,\qquad t>T,
\end{equation}
where $T$ and $M$ are some constants. To achieve the boundary condition,
one would use a contour which does not enter the complex right half plane.
The solution would then be unique, if and only if the only singularity of
the integrand in the right hand side of (\ref{02.14}) in the complex left
half plane is a simple pole. That is the case if in the complex left
half plane, $f$ has only poles (no other singularities) and just one simple
zero.
\section{Moments corresponding to a distribution}
Consider a density $\psi$ defined on the space. The moments
corresponding to such a distribution are expectation values of
monomials of space coordinates:
\begin{align}\label{02.19}
\M(\bm{\alpha}):=&\,\left\langle\prod_j(r^j)^{\alpha_j}\right\rangle,\nonumber\\
=&\,\left[\int\d V\;\psi(\bm{r})\right]^{-1}\,
\left[\int\d V\;\psi(\bm{r})\,\prod_j(r^j)^{\alpha_j}\right],
\end{align}
where $(r^1,\dots)$ are the coordinates of $\bm{r}$.
Taking the space to be $\mathbb{R}^n$, these can be expressed in terms of
the Fourier transform of the density. Denoting the Fourier transform of $\psi$
by $(\F\psi)$,
\begin{equation}\label{02.20}
(\F\psi)(\bm{k}):=\int\d V\;\exp(-\ri\,\bm{k}\cdot\bm{r})\,\psi(\bm{r}),
\end{equation}
one has
\begin{equation}\label{02.21}
\M(\bm{\alpha})=[(\F\psi)(\bm{0})]^{-1}\,
\left[\prod_j\left(\ri\,\frac{\partial}{\partial k_j}\right)^{\alpha_j}(\F\psi)\right](\bm{0}).
\end{equation}
The connected moments are defined as the derivatives of the logarithm of the
Fourier transform of the density:
\begin{equation}\label{02.22}
\M_\mathrm{c}(\bm{\alpha}):=
\left\{\prod_j\left(\ri\,\frac{\partial}{\partial k_j}\right)^{\alpha_j}[\ln(\F\psi)]\right\}(\bm{0}).
\end{equation}
Examples of the connected moments are the expectation values of coordinates:
\begin{equation}\label{02.23}
\langle r^j\rangle=\left\{\ri\,\frac{\partial}{\partial k_j}[\ln(\F\psi)]\right\}(\bm{0}),
\end{equation}
and correlations of coordinates:
\begin{equation}\label{02.24}
\left\langle(r^j-\langle r^j\rangle)\,(r^l-\langle r^l\rangle)\right\rangle=
\left\{\left(\ri\,\frac{\partial}{\partial k_j}\right)
\left(\ri\,\frac{\partial}{\partial k_l}\right)
[\ln(\F\psi)]\right\}(\bm{0}),
\end{equation}
from which one constructs the variance:
\begin{align}\label{02.25}
\Var(\bm{r}):=&\,\left\langle(\bm{r}-\langle\bm{r}\rangle)\cdot(\bm{r}-\langle\bm{r}\rangle)\right\rangle,
\nonumber\\
=&\,-\left\{\nabla_{\bm{k}}^2[\ln(\F\psi)]\right\}(\bm{0}).
\end{align}
\section{Time evolution of distributions}
Consider a distribution $\psi$ satisfying the equation
\begin{equation}\label{02.26}
[f(\D_0,\Df)]\psi=0,
\end{equation}
where $\Df$ is differentiation with respect to $\bm{r}$ (the space variable).
Space (time) translational invariance means that $f$ does not depend on
the space (time) variable. It is assumed that this is the case. 
Note that such operators include any space operator subject to 
only translational invariance, including fractional space
derivatives with the effect that the space Fourier 
transform is multiplied by a function of the Fourier variable. 

The Fourier transform of (\ref{02.26}) reads
\begin{equation}\label{02.27}
[f(\D_0,\ri\,\bm{k})](\F\psi)=0.
\end{equation}
The criterion that the solution to this which does not blow up
at $t\to\infty$ be unique (knowing only one initial condition) is
that in the complex left hand plane $f(\centerdot,\ri\,\bm{k})$ has
only poles and just one simple zero. Denoting that zero by $E(\bm{k})$,
one arrives at
\begin{equation}\label{02.28}
(\F\psi)(t,\bm{k})=[(\F\psi)(0,\bm{k})]\,\exp[t\,E(\bm{k})].
\end{equation}
Then, using (\ref{02.22}) it turns out that
\begin{align}\label{02.29}
[\M_\mathrm{c}(\bm{\alpha})](t)&=
\left\{\prod_j\left(\ri\,\frac{\partial}{\partial k_j}\right)^{\alpha_j}[\ln(\F\psi)]\right\}(0,\bm{0})
+t\,\left[\prod_j\left(\ri\,\frac{\partial}{\partial k_j}\right)^{\alpha_j}E\right](\bm{0}),\nonumber\\
&=[\M_\mathrm{c}(\bm{\alpha})](0)
+t\,\left[\prod_j\left(\ri\,\frac{\partial}{\partial k_j}\right)^{\alpha_j}E\right](\bm{0}).
\end{align}
This shows that the connected moments are at most linear functions of time.
Specifically, the variance changes linearly with time. These conclusions
are valid unless the corresponding derivatives of $E$ at $\bm{k}=\bm{0}$
either are zero or blow up. The moment would be constant in the former case,
and would be infinite in the latter. So no fractional power time dependence
would arise.

A similar conclusion holds for general (not necessarily connected) moments.
Using (\ref{02.21}) and (\ref{02.28}), it is seen that
\begin{equation}\label{02.30}
[\M(\bm{\alpha})](t)=[\M(\bm{\alpha})](0)+\mathrm{P}_{\bm{\alpha}}(t),
\end{equation}
where $\mathrm{P}_{\bm{\alpha}}$ is a polynomial of degree at most
$\mathrm{deg}(\bm{\alpha})$ in $t$, with
\begin{equation}\label{02.31}
\mathrm{deg}(\bm{\alpha}):=\sum_j\alpha_j.
\end{equation}
Exceptions are again when the derivatives of $E$ at $\bm{k}=\bm{0}$ blow up, but
even in that case one does not encounter a fractional power law evolution either.

As an example of a linear fractional time derivative equation, which is invariant
under time and space translations, consider
\begin{equation}\label{02.32}
(-1)^{[\beta]}\,\sideset{^\mathrm{W}}{_{0\,\beta}}{\mathop{\D}}\,\psi+\Df\cdot\Df\,\psi=0,
\end{equation}
where the branch cut for the power function with the exponent $\beta$, is the positive
real semi axis. Using (\ref{02.11}), one has
\begin{equation}\label{02.33}
(-1)^{[\beta]}\,\exp[-\ri\,\pi\,\r(\beta)]\,[E(\bm{k})]^\beta-\bm{k}\cdot\bm{k}=0,
\end{equation}
or
\begin{equation}\label{02.34}
[\exp(-\ri\,\pi)\,E(\bm{k})]^\beta=\bm{k}\cdot\bm{k}.
\end{equation}
Noting that the angle of $E(\bm{k})$ is between 0 and $(2\,\pi)$, it is seen that
one solution of this equation is
\begin{equation}\label{02.35}
E(\bm{k})=-(\bm{k}\cdot\bm{k})^{1/\beta}.
\end{equation}
The solution is unique, provided $\beta$ is less than 2. Assuming that's the case,
one arrives at
\begin{equation}\label{02.36}
(\F\psi)(t,\bm{k})=[(\F\psi)(0,\bm{k})]\,\exp[-t\,(\bm{k}\cdot\bm{k})^{1/\beta}],
\end{equation}
which shows that there is no power law time dependence with fractional exponents,
for connected moments, although such moments could be constant or blow up, when
$\beta$ is not equal to one.
\section{Lack of time translation symmetry, and nonconstant diffusion rate}
As an example of a fractional time derivative evolution which results in a nonconstant
diffusion rate, consider
\begin{equation}\label{02.37}
\sideset{_0^\mathrm{C}}{_{0\,\beta}}{\mathop{\D}}\,\psi=\Df\cdot\Df\,\psi,
\end{equation}
where
\begin{equation}\label{02.38}
\beta=:1-\epsilon,
\end{equation}
and
$\epsilon$ is a small positive number. Using (\ref{02.7}), one arrives at
\begin{align}\label{02.39}
(\sideset{_0^\mathrm{C}}{_{0\,\beta}}{\mathop{\D}}\,\psi)(t,\bm{r})
&=\frac{1}{\Gamma(1+\epsilon)}\,\left[t^\epsilon\,(\D_0\psi)(t,\bm{r})
+\int_0^t\d\tau\;(t-\tau)^\epsilon\,(\D_0^2\psi)(\tau,\bm{r})\right],\nonumber\\
&=(\D_0\psi)(t)+\epsilon\,\gamma\,(\D_0\psi)(t)\nonumber\\
&\quad+\epsilon\,\left[(\D_0\psi)(0)\,\ln t+
\int_0^t\d\tau\;\ln(t-\tau)\,(\D_0^2\psi)(\tau)\right]+o(\epsilon),
\end{align}
where $\gamma$ is the Euler constant. To obtain a perturbative
(in $\epsilon$) solution of (\ref{02.37}), one considers
the three last terms on the right hand side of (\ref{02.39})
as sources applied to an ordinary diffusion equation. Then
the Fourier transform of (\ref{02.37}) reads
\begin{equation}\label{02.40}
(\D_0\F\psi)(t,\bm{k})+J(t,\bm{k})=-\bm{k}\cdot\bm{k}\,\F\psi(t,\bm{k}),
\end{equation}
where
\begin{equation}\label{02.41}
J(t,\bm{k})=J_1(t,\bm{k})+J_2(t,\bm{k})+o(\epsilon),
\end{equation}
and
\begin{align}\label{02.42}
J_1(t,\bm{k})&=\epsilon\,\gamma\,(\D_0\F\psi^\mathrm{H})(t),\nonumber\\
J_2(t,\bm{k})&=\epsilon\,\left[(\D_0\F\psi^\mathrm{H})(0)\,\ln t+
\int_0^t\d\tau\;\ln(t-\tau)\,(\D_0^2\F\psi^\mathrm{H})(\tau)\right],
\end{align}
where $\psi^\mathrm{H}$ is the solution to (\ref{02.40}) without
the source ($J$) term. It is seen that the relation of $J_1$ to $\psi$
is time translation invariant, while that of $J_2$ with $\psi$ is not.

The solution to (\ref{02.40}) is
\begin{equation}\label{02.43}
(\F\psi)(t,\bm{k})=(\F\psi^\mathrm{H})(t,\bm{k})+
\int_0^\infty\d t'\;G(t-t',\bm{k})\,J(t',\bm{k}),
\end{equation}
where
\begin{equation}\label{02.44}
(\F\psi^\mathrm{H})(t,\bm{k})=(\F\psi)(0,\bm{k})\,\exp(-\bm{k}\cdot\bm{k}\,t),
\end{equation}
and $G$ is the Green's function, satisfying
\begin{align}\label{02.45}
(\D_0G)(t,\bm{k})+\delta(t)&=-\bm{k}\cdot\bm{k}\,G(t,\bm{k}),\nonumber\\
G(0^-,\bm{k})&=0.
\end{align}
One arrives at
\begin{equation}\label{02.46}
G(t,\bm{k})=-\theta(t)\,\exp(-\bm{k}\cdot\bm{k}\,t),
\end{equation}
where $\theta(t)$ is the Heaviside step-function. One then has,
\begin{align}\label{02.47}
(\F\psi_1)(t,\bm{k})&=\epsilon\,\gamma\,\bm{k}\cdot\bm{k}\,
\int_0^t\d t'\;\exp[-\bm{k}\cdot\bm{k}\,(t-t')]\,(\F\psi^\mathrm{H})(t',\bm{k}),\\ \label{02.48}
(\F\psi_2)(t,\bm{k})&=\epsilon\,\gamma\,\bm{k}\cdot\bm{k}\,
(\F\psi^\mathrm{H})(0,\bm{k})\,\int_0^t\d t'\;\exp[-\bm{k}\cdot\bm{k}\,(t-t')]\,\ln t'\nonumber\\
&\quad-\epsilon\,(\bm{k}\cdot\bm{k})^2\,\int_0^t\d t'\;\exp[-\bm{k}\cdot\bm{k}\,(t-t')]\nonumber\\
&\qquad\times\int_0^{t'}\d\tau\,\ln(t'-\tau)\,(\F\psi^\mathrm{H})(t',\bm{k}),
\end{align}
where
\begin{equation}\label{02.49}
(\F\psi_i)(t,\bm{k}):=\int_0^\infty\d t'\;G(t-t',\bm{k})\,J_i(t',\bm{k}).
\end{equation}
Finally, using (\ref{02.25}) one arrives at
\begin{equation}\label{02.50}
[\Var(\bm{r})](t)=[\Var(\bm{r})](0)+2\,d\,t-2\,d\,\epsilon\,\gamma\,\left(t+\int_0^t\d t'\,\ln t'\right)
+o(\epsilon),
\end{equation}
where $d$ is the dimension of the space. It is seen that it is only the
integral of the logarithm in the right hand side, which results in
a fractional power of time appearing in the variance:
\begin{align}\label{02.51}
[\Var(\bm{r})](t)&=[\Var(\bm{r})](0)+\frac{2\,d\,t^{1-\epsilon}}{(1-\epsilon)\,\Gamma(1-\epsilon)}
+o(\epsilon),\nonumber\\
&=[\Var(\bm{r})](0)+\frac{2\,d\,t^\beta}{\Gamma(1+\beta)}.
\end{align}
The last line is in fact exact, as can be obtained by the exact solution of
(\ref{02.37}). The integral of the logarithm in the right hand side of (\ref{02.50})
comes from $J_2$, which has a time translation noninvariant relation to $\psi$. So the
origin of the fractional power of time in the variance is lack of
the time translation invariance of the evolution equation.
\section{Concluding remarks}
It was shown that a system satisfying the conditions
of linearity, both space and time translational invariance, and the property
that the state of the system at one time determines uniquely the state of
the system at later times, does not show anomalous diffusion.
If these criteria are met, then the time dependence of the connected moments
are at most linear functions of time, while the time dependence of
a general moment of order $\alpha$ is a polynomial of degree not larger
than $\alpha$. This result holds regardless of the specific operators
entering the evolution of system, even if they contain fractional
time or space derivatives, provided of course the above criteria are met.
Two examples of fractional time derivative evolutions were also explicitly
studied. One of them (the Weyl derivative) satisfies the above conditions,
hence does not exhibit anomalous diffusion. The other (the Caputo) does not
satisfy the above conditions, and does exhibit anomalous diffusion. In the latter
case, the anomalous behavior was traced back to the term in the Caputo fractional
derivative which violates time translational invariance.
\\[\baselineskip]
\textbf{Acknowledgement}: This work was partially
supported by the Research Council of the Alzahra University.


\newpage
\begin{thebibliography}{99}
\bibitem{meka2000}   R. Metzler \& J. Kalfter, Phys. Rep. \textbf{339} (2000) 1-77.

\bibitem{hughes}     B. D. Hughes, ``Random Walks and Random Environments, Vol. 1: Random Walks''
                     (Oxford University Press, 1995)

\bibitem{bouch}      J. P. Bouchaud \& A. Georges, Phys. Rep. \textbf{195} (1990) 127

\bibitem{meka2004}   R. Metzler \& J. Klafter, J. Phys. \textbf{A37} (2004) 1505.

\bibitem{vlahos}     L. Vlahos, H. Isliker, Y. Kominis, \& K. Hizanidis, arXiv:0805.0419.

\bibitem{blu1986}    A. Blumen A, J. Klafter, \& G. Zumofen, in ``Optical Spectroscopy of Glasses''
                     I. Zschokke (ed.) (Reidel, 1986)

\bibitem{shles}      M. F. Shlesinger, G. M. Zaslavsky, \& J. Klafter, Nature \textbf{363} (1993) 31

\bibitem{barkai2001} E. Barkai, Phys. Rev. \textbf{E63} (2001) 046118

\bibitem{hilfer2000} R. Hilfer (ed.), Applications of Fractional Calculus in Physics
                     (World Scientific, 2000)

\bibitem{meka2001}   R. Metzler \& J. Klafter, Adv. Chem. Phys. \textbf{116} (2001) 223

\bibitem{escande}    D. F. Escande \& F. Sattin, Phys. Rev. Lett. \textbf{99} (2007) 185005.

\bibitem{kl1987}     J. Klafter, A. Blumen, \& M. F. Shlesinger, Phys. Rev. \textbf{A35} (1987) 3081.

\bibitem{lenzi}      E. K. Lenzi, R. S. Mendes, \& C. Tsallis, Phys. Rev. \textbf{E67} (2003) 031104.

\bibitem{frac1}      A. A. Kilbas, H. M. Srivastava, \& J. J. Trujillo, ``Theory and applications of
                     fractional differential equations'', (Elsevier 2006).

\bibitem{frac2}      I. Podlubny, ``Fractional differential equations'', (Academic Press 1999).
\end{thebibliography}
\end{document}